\documentclass[aps,prl,reprint,nofootinbib,superscript address]{revtex4-1}
\usepackage{amssymb}

\usepackage{tabularx}
\usepackage{subfigure}
\usepackage{amsmath,amssymb}
\usepackage{latexsym}
\usepackage{psfrag}
\usepackage{epsfig}
\usepackage{graphicx,subfigure}
\usepackage{array}
\usepackage{tabularx}

\usepackage{multirow}
\usepackage{epstopdf}
\usepackage{graphicx}

\usepackage{algorithm}

\begin{document}
\title{On-line Quantum State Estimation Using Continuous Weak Measurement and Compressed Sensing}
\author{S. Cong}
\email{scong@ustc.edu.cn}
\affiliation{Department of Automation, University of Science and Technology of China, Hefei 230027, P. R. China}
\author{Y. Tang}
\affiliation{Department of Automation, University of Science and Technology of China, Hefei 230027, P. R. China}
\author{S. Harraz}
\affiliation{Department of Automation, University of Science and Technology of China, Hefei 230027, P. R. China}
\author{K. Li}
\email{kezhi.li@imperial.ac.uk}
\affiliation{Department of Electrical and Electronic Engineering, Imperial College London, SW7 2AZ, UK}
\author{J. Yang}
\affiliation{Department of Automation, University of Science and Technology of China, Hefei 230027, P. R. China}

\begin{abstract}
We propose a new protocol for on-line quantum system estimation on the basis of continuous weak-measurements with the help of compressive sensing and the optimization algorithm. By directly measuring the state of the probe system, we indirectly obtain the measurement operators of the estimated system. The continuous weak measurements for a dynamic evolution open quantum system enable us to derive the time-varying measurement operators of the estimated system. Compressed sensing is used to reduce the number of the measurements needed and to improve the efficiency of the estimation. This approach to the on-line state estimation provides a novel solution to the problem of closed-loop quantum feedback control.
\end{abstract}

\maketitle

Quantum state estimation (QSE), also called Quantum state tomography (QST), is usually formulated by means of strong (projective) measurements of an informationally complete set of measurement operators and corresponding observables [1]. However, strong measurements collapse the original quantum state, the ensemble must be reprepared and the measurement apparatus has to be reconfigured at each step. Weak measurements (WM) [2] offer an alternative in acquiring quantum measurements and estimating quantum states. In the measuring process, by using continuous weak measurements (CWM) it is possible to gain the target state information without disturbing it substantially, and the value recovered in CWM can be obtained by computing the ensemble averaging. The first protocol for the continuous measurement of QST was proposed by Silberfarb [3] and was implemented in [4], [5]. Usually, the number of measurements required for a
 \textit{d}-dimension density matrix estimation of an $n$-qubit system is specified by $O(d\times d)=O(2^{n} \times 2^{n} )=O(4^{n} )$ [6], which grows exponentially with the system size. Compressed Sensing (CS) [7] has been brought into the quantum domain in the context of reducing the number of measurements required for QST [8], [9],[10], Quantum Process Tomography [11], measurement of complementary observables [12] and the quantum wavefunction [13]. However, whether a unified efficient scheme  for on-line quantum state estimation using CWM and partial measurements is feasible, remains unknown. In this letter, we propose a new protocol for on-line quantum state estimation on the basis of continuous weak-measurements with the help of compressive sensing and the optimization algorithm. Our key idea is to make a weak measurement on the complete measurement operators in an ensemble system by coupling the ensemble to some probe which can be measured. At each instant time, we obtain the records of the expectation values corresponding to some measurement operators using the indirect results of continuous weak measurements, and the estimated state is obtained by solving an on-line optimization problem with physical constraints. CS is used to reduce the number of the measurements needed and to improve the efficiency of the estimation. In our scheme, CWM and the state estimation are carried out on-line continuously, while the existing techniques usually perform continuous weak measuring on-line yet estimate the state off-line, and known that the estimated state is a fixed state. During the on-line state estimation in this letter, the state of quantum system is in a dynamic evolution, and the states estimated on-line are time-dependent states of the system, so the measurement operators in the on-line state estimation are no longer a fixed matrix group. They become a set of time varying measurement operators. The on-line state estimation really solves the problem of closed-loop quantum feedback control. Our procedure is broadly applicable in systems where continuous weak measurement tools have been developed, such as nuclear magnetic resonance in molecules [14] and polarization spectroscopy in atomic vapors [15].

The process of quantum weak measurement is shown in Fig. 1, which consists of two parts: detection part and readout part. A probe $P$ is coupled with the estimated system $S$, and they become a joint coupled system $S\otimes P$. For one qubit density matrix $\rho $, suppose the initial state of the probe $P$ is ${\left| \phi  \right\rangle} $, and the initial state of the system $S$ is $\rho _{0} ={\left| \psi  \right\rangle} {\left\langle \psi  \right|} $. $H_{S} $ and $H_{P} $ are the Hamiltonians of system $S$ and $P$, respectively, and $H=H_{P} \otimes H_{S} $ is the Hamiltonian of the joint system.

\begin{figure}[!htbp]
\centering
\includegraphics[height=0.15\textwidth, 
width=0.45\textwidth]{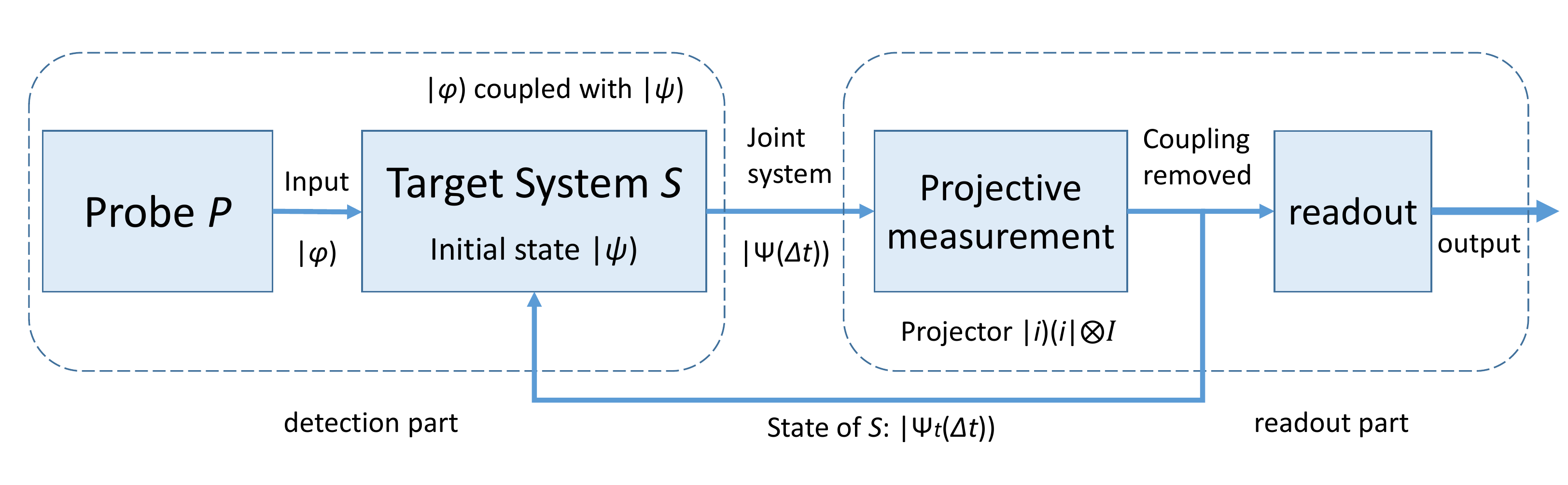}
\caption{Process of quantum weak measurement}
\label{fig1}
\end{figure}

The initial state of the coupled system is ${\left| \Psi  \right\rangle} $: ${\left| \Psi  \right\rangle} ={\left| \phi  \right\rangle} \otimes {\left| \psi  \right\rangle} $. After the joint evolution of $S$ and $P$ for time $\Delta t$, the state ${\left| \Psi  \right\rangle} $ becomes ${\left| \Psi (\Delta t) \right\rangle} =U(\Delta t){\left| \Psi  \right\rangle} $, where $U(\Delta t)$ is the joint evolution operator $U(\Delta t)=\exp (-i\xi \Delta tH/\hbar )$, and $\xi $ represents the interaction strength between system $S$ and $P$. ${\left| \Psi (\Delta t) \right\rangle} $ is an entangled state composed of $S$ and $P$, which cannot be separately described with the states of $S$ and $P$. At time $\Delta t$, a measurement is performed on the entangled state with the measurement operator $X=\sum I\otimes {\left| k \right\rangle} {\left\langle k \right|}  $, where ${\left| k \right\rangle} $ is the eigenstate of the system $P$: ${\left| 0 \right\rangle} $ or ${\left| 1 \right\rangle} $ . This measurement is actually a projective measurement on $P$, and the outputs are the eigenvalues corresponding to ${\left| k \right\rangle} $. The state of the joint system after the weak measurement becomes
\begin{equation} \label{eq1}
{\left| \Psi _{k} (\Delta t) \right\rangle} {\rm =(}{\left| k \right\rangle} {\left\langle k \right|} \otimes I\cdot U(\Delta t){\left| \phi  \right\rangle} \otimes {\left| \psi  \right\rangle} {\rm )/}\Theta _{k},
\end{equation}
where $\Theta _{k} {\rm =}\sqrt{{\left\langle \Psi (\Delta t) \right|} \Pi _{k} {\left| \Psi (\Delta t) \right\rangle} } $.

After the projective measurement, the entanglement between $S$ and $P$ disappears, and the state of \textit{S} corresponds to the output at time $\Delta t$ becomes ${\left| \psi _{k} (\Delta t) \right\rangle} $. The state of the joint system after the weak measurement can also be represented as
\begin{equation} \label{eq2}
{\left| \Psi _{k} (\Delta t) \right\rangle} {\rm =}{\left| k \right\rangle} \otimes {\left| \psi _{k} (\Delta t) \right\rangle}.
\end{equation}
Putting Eq. (2) into the Eq. (2), we can obtain
\begin{equation} \label{eq3}
{\left| \psi _{k} (\Delta t) \right\rangle} {\rm =}{\left\langle k \right|} \otimes I\cdot U(\Delta t){\left| \phi  \right\rangle} \otimes {\left| \psi  \right\rangle} /\Theta _{k}.
\end{equation}
We define the weak measurement operator $M_{k} $ as
\begin{equation} \label{eq4}
M_{k} ={\left\langle k \right|} \otimes I\cdot U(\Delta t)\cdot {\left| \phi  \right\rangle} \otimes I,
\end{equation}
which is a Kraus operator and satisfies $\sum _{k}M_{k}^{\dag } M_{k}  =1$. In this case, $\Theta _{k} $ becomes
\begin{equation} \label{eq5}
\Theta _{k} {\rm =}\sqrt{{\left\langle \psi  \right|} M_{k} {}^{\dag } M_{k} {\left| \psi  \right\rangle} }.
\end{equation}
By substituting Eqs. (4) and (5) in (3), we can get the relationship between the state of the system $S$ before and after the whole measurement process as:
\begin{equation} \label{eq6}
{\left| \psi _{k} (\Delta t) \right\rangle} {\rm =}\frac{M_{k} }{\sqrt{{\left\langle \psi  \right|} M_{k} {}^{\dag } M_{k} {\left| \psi  \right\rangle} } } {\left| \psi  \right\rangle}.
\end{equation}

In such a way, we obtain the weak measurement operator $M_{k} $ in Eq. (4) on the system $S$.

On-line state estimation makes the measurement operators be no longer a constant matrix group, and they become a set of time varying measurement operators $M_{k} (t)$. We need to deduce the time varying measurement operators used in the on-line state estimation.

\begin{figure}[!htbp]
\centering
\includegraphics[height=0.15\textwidth, width=0.45\textwidth]{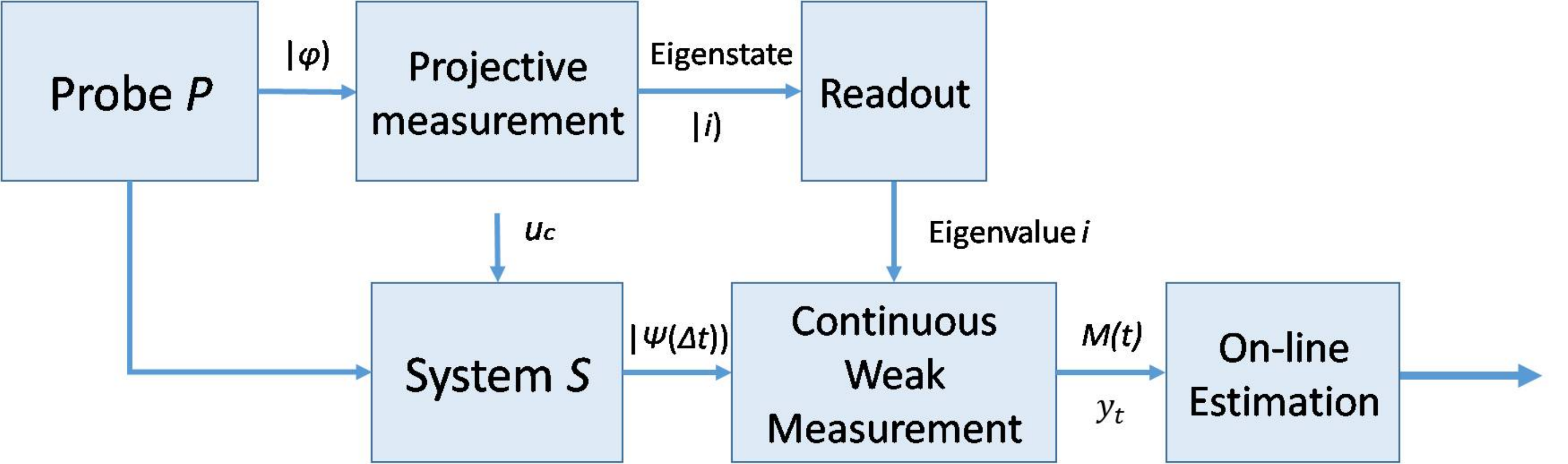}
\caption{ On-line estimation of quantum state based on continuous weak measurements}
\label{fig2}
\end{figure}

The process of on-line estimation of quantum states based on continuous weak measurements is shown in Fig. 2. Let $\lambda =\xi \Delta t$ denote the weak measurement strength, where both the interaction strength $\xi $ and the evolution time $\Delta t$ are small values. Assume $\hbar =1$, weak measurement is in the case of $\xi \Delta t\to 0$. We can get the Taylor expansion of $U$ and neglect more than three orders of magnitude as $U(\Delta t)\approx I\otimes I-i\xi \Delta tH-{(\xi \Delta t)^{2} H^{2} \mathord{\left/ {\vphantom {(\xi \Delta t)^{2} H^{2}  2}} \right. \kern-\nulldelimiterspace} 2} $, which is put into Eq. [4]. At the same time, the Taylor expansion is carried out, and the minimum amount of more than three orders is ignored, we can obtain the expression of the weak measurement operator as $M_{k} (\Delta t)\approx I\left\langle \left. k\right|\right. \left. \phi \right\rangle -i\xi \Delta tH_{S} \left\langle \left. k\right|\right. H_{P} \left|\left. \phi \right\rangle \right. -{(\xi \Delta t)^{2} H_{S}^{2} \left\langle \left. k\right|\right. H_{P}^{2} \left|\left. \phi \right\rangle \right. \mathord{\left/ {\vphantom {(\xi \Delta t)^{2} H_{S}^{2} \left\langle \left. k\right|\right. H_{P}^{2} \left|\left. \phi \right\rangle \right.  2}} \right. \kern-\nulldelimiterspace} 2} $.  Let $r_{k} ={(\xi \Delta t)H_{S}^{2} \left\langle \left. k\right|\right. H_{P}^{2} \left|\left. \phi \right\rangle \right. \mathord{\left/ {\vphantom {(\xi \Delta t)H_{S}^{2} \left\langle \left. k\right|\right. H_{P}^{2} \left|\left. \phi \right\rangle \right.  2}} \right. \kern-\nulldelimiterspace} 2} , k=1,2,...,d$ , and the general form of the weak measure operator is: $M_{k} (\Delta t)=I\left\langle \left. k\right|\right. \left. \phi \right\rangle -[r_{k} \lambda /2+i\lambda H_{S} \left\langle \left. k\right|\right. H_{P} \left|\left. \phi \right\rangle \right. ]$.  Suppose ${\left\langle j \mathrel{\left| \vphantom{j \phi }\right.\kern-\nulldelimiterspace} \phi  \right\rangle} =1$ when $k=j$, we can obtain $M_{i} (t)$ as : $M_{j} (\Delta t)=I-(\xi r_{k=j} /2+i\xi H_{S} )\Delta t$, and all the other measurement operators of $k\ne j$ can be combined as one operator as: $M_{k\ne j} (\Delta t)=M_{j\bot } (\Delta t)=\sqrt{r_{k\ne j} \Delta t} $, where  $M_{j\bot } $ and $M_{j} $ are orthogonal and satisfy $(M_{j\bot } {\rm )}^{{\rm 2}} {\rm +}(M_{j} {\rm )}^{{\rm 2}} =I$. For the continuous weak measurements of a two-level quantum systems, the measurement operator group only contains two operators: $M_{0} (\Delta t)$ and $M_{1} (\Delta t)$. By selecting the appropriate operator $L$, the corresponding continuous weak measurement operators $M_{0} (\Delta t)$ and $M_{1} (\Delta t)$ can be constructed, respectively, as
\begin{equation} \label{eq7}
\begin{aligned}
 M_{0} (\Delta t) &=M_{j} -i(1-\xi )H_{S} \Delta t \\ &=I-\left({\xi r_{k} \mathord{\left/ {\vphantom {\xi r_{k}  2}} \right. \kern-\nulldelimiterspace} 2} +iH(t)\right)\Delta t \\  &=I-\left({L^{\dag } L\mathord{\left/ {\vphantom {L^{\dag } L 2}} \right. \kern-\nulldelimiterspace} 2} +iH(t)\right)\Delta t, \\ M_{1} (\Delta t)&=M_{k\ne j} =L\cdot \sqrt{\Delta t},
 \end{aligned}
\end{equation}
in which $L^{\dag } L=\xi r_{k} $.

The stochastic master equation (SME) of the open quantum system can be written as:
\begin{equation} \label{eq8}
\begin{aligned}
&{\rho (t+dt)-\rho (t)}= {-\frac{i}{\hbar } [H(t),\rho (t)]dt}+ \\& \sum _{}\left[L\rho (t)L^{\dag } -\left(\frac{1}{2} L^{\dag } L\rho (t)+\frac{1}{2} \rho (t)L^{\dag } L\right)\right] dt \\ {} &{+\sqrt{\eta } \sum _{}\left[L\rho (t)+\rho (t)L^{\dag } \right] dW}, \\ &{\rho _{0} } = {\rho (0)},
\end{aligned}
\end{equation}
in which $\rho (t)$ is the density matrix; $H(t)=H_{S} +H_{P} +u(t)H_{c} $, $H(t)$ is the whole Hamiltonian; $H_{S} $ is the measured system Hamiltonian, $H_{P} $ is the Hamiltonian of Probe system; $H_{c}$ is the control Hamiltonian. $u(t)$ is the external regulate value; $\eta$ is the measure efficiency and satisfies $0<\eta \le 1$; Let \textbf{$D[L,\rho ]=L\rho L^{\dag } -(1/2)\left(L^{\dag } L\rho +\rho L^{\dag } L\right)$}, which is the decoherence effect of the measurement process, and a drift term expressed as a Lindblad form; let $H[L,\rho ]=L\rho +\rho L^{\dag } $, the stochastic diffusion term introduced by the measurement process is expressed as a disturbance to the state of the quantum system, also known as the reverse effect (back-action). In the condition of homodyne measurement, the noise produced by measurement output for zero error measurement is a one-dimensional Wiener process, and it satisfies $E(dW)=0,E[(dW)^{2} ]=dt$. From the continuous weak measurement process of the quantum system, we can see that the measurement process actually contains the system evolution, so the continuous weak measurement operator $M_{0} (\Delta t)$ contains the system total Hamiltonian $H(t)$. If the system random noise and measurement efficiency are both considered in the measurement process, the evolution operators of the system become
\begin{equation} \label{eq9}
\begin{array}{rcl} {A_{0} } & {=} & {M_{0} (dt)+\sqrt{\eta } L\cdot dW}, \\ {A_{1} } & {=} & {M_{1} (dt)+\sqrt{\eta } L\cdot dW}, \end{array}
\end{equation}
where $dt=\Delta t$ represents the very short time interval required for the weak measurement, $L\cdot dW$ denotes the noise caused by the continuous weak measurements, $dW$ denotes Gaussian white noise, and $W(t)$ is a Weiner process with zero mean $E\left[W(t)\right]=0$ and unit variance $E[\left(W(t)\right)^{2} ]=dt$.

The discrete-time dynamic evolution equation of the stochastic open quantum system $S$ can be written as:
\begin{equation} \label{eq10}
\rho (t+dt)=A_{0} \rho (t)A_{0}^{\dag } +A_{1} \rho (t)A_{1}^{\dag }.
\end{equation}
\noindent The evolution equation of the measurement operator $M_{i} (t)$ is: $\dot{M}_{k} (t)=\frac{i}{\hbar } [H(t),M_{k} (t)]+LM_{k} (t)L^{\dag } -\frac{1}{2} \left(L^{\dag } LM_{k} (t)+M_{k} (t)L^{\dag } L\right)$, the corresponding discrete-time evolution equation of continuous weak measurement operators are:
\begin{equation} \label{eq11}
M_{k} (t+dt)=M_{0} {}^{\dag } M_{k} (t)M_{0} +M_{1} {}^{\dag } M_{k} (t)M_{1}
\end{equation}


According to the theory of compressed sensing (CS), the density matrix of quantum state can be reconstructed with only $O\left(rd\ln d\right)$ measurements' numbers of random measurement operators by solving an optimization problem, where \textit{r} and \textit{d }are the dimension and rank of the density matrix $\rho $, respectively, and $r\ll d$ [16]. Here we use the following estimator of minimizing the 2-norm under the positive definite constraint:
\begin{equation} \label{eq12}
\begin{array}{l} {\hat{\rho }=\arg \min \left\| {\bf A}\  \cdot vec(\rho )-y\right\| _{2} {\rm \; \; }} \\ {{\rm s.t.\; \; \; }\hat{\rho }=\rho ,{\rm \; }\rho \ge 0,{\rm tr(}\rho )=1} \end{array},
\end{equation}
where $vec(\cdot )$ represents the transformation from a matrix to a vector by stacking the matrix's columns in order on the top of one another. The sampling matrix $\bf A$ is the matrix form of the all the sampled measurement operators $M_{k_{l} } (t_{l} )$; $M_{k_{l} } $, $l=1,2,...,m$ is an arbitrary measurement operator in the \textit{l}-th or the $t_{l} $-th measurement. For the sake of simplicity, we let $M_{k_{l} } =M_{k_{l} } (t_{l} )$. The vector $y$ and matrix $\bf A$ can be expressed according to the current measurement configurations as [17]:
\begin{equation} \label{eq13}
y(t_{l} )=(\left\langle M_{k_{1} } {\kern 1pt} \right\rangle ,\left\langle M_{k_{2} } \right\rangle ,\cdot \cdot \cdot ,\left\langle M_{k_{m} } \right\rangle )^{T} , l=1,2,...,m,
\end{equation}
and
\begin{equation}\label{eq14}
\begin{aligned}
{\bf A}  (t_{l} )&=\left(\begin{array}{cccc} {vec(M_{k_{1} } )^{T} } & {vec(M_{k_{2} } )^{T} } & {\cdots } & {vec(M_{k_{m} } )^{T} } \end{array}\right), \\ l&=1,2,...,m,
\end{aligned}
\end{equation}

\noindent where $\left\langle M_{k_{i} } \right\rangle $, $l=1,2,...,m$ is the corresponding measurement value in the \textit{l}-th measurement; the sampling vector $y$ is the vector form of the corresponding observation values $\left\langle M_{k_{l} } \right\rangle $.

One can estimate the quantum state on-line with a small amount of time-evolving measurement operators $\left\{M_{k_{l} } \right\}$, $l=1,2,...,m$ and corresponding measures records $y(t_{l} )$, $l=1,2,...,m$ by solving the optimization problem (12) with an appropriate algorithm.

Consider a 1/2 spin particle ensemble $\rho (t)$ as the system of on-line state estimation, which is under $z$ direction with a constant magnetic field $B_{z} $ and $x$ direction control magnetic field $B_{x} =A\cos \phi $. In Schr\"{o}dinger picture, the initial state of the spin is $\rho (0)$, and $\rho (t)$ represents the state at moment $t$. The eigen-frequency of the spin ensemble $\rho (t)$ in the magnetic field $B_{z} $ is $\omega _{0} =\gamma B_{z} $, where $\gamma $ is the spin-magnetic ratio of the particle ensemble, and $\Omega =\gamma A$ is the Rabi frequency of the system $\Omega \in {\rm R}$. The Hamiltonian of system is: $H=H_{0} +u_{x} H_{x} $, where $H_{0} =-\left({\hbar \mathord{\left/ {\vphantom {\hbar  2}} \right. \kern-\nulldelimiterspace} 2} \right)\omega _{0} \sigma _{z} $ is the free Hamiltonian, $\sigma _{z} =\left[\begin{array}{cc} {1} & {0} \\ {0} & {-1} \end{array}\right]$ is the Pauli operator of \textit{z}, $u_{x} H_{x} =-\hbar \omega _{0} {\left(e^{-i\phi } \sigma ^{-} +e^{i\phi } \sigma ^{+} \right)\mathord{\left/ {\vphantom {\left(e^{-i\phi } \sigma ^{-} +e^{i\phi } \sigma ^{+} \right) 2}} \right. \kern-\nulldelimiterspace} 2} =-\gamma A\sigma _{x} $ is the control Hamiltonian; $\sigma ^{-} =\left[\begin{array}{cc} {0} & {0} \\ {1} & {0} \end{array}\right]$, $\sigma ^{+} =\left[\begin{array}{cc} {0} & {1} \\ {0} & {0} \end{array}\right]$, and $u_{x} \in {\rm R}^{+} $ is the time-independent control amplitude: $u_{x} =\gamma A$. The initial phase of control field is $\phi =0$. A continuous weak measurement is applied to the system. The initial weak measurement operator is $M_{k} $ in Eq. [4]. The sampling matrix $\bf A$ and $y(t_{l} )$ are calculated according to Eq. [13] and Eq. [14]. We use the least-square algorithm to solve the optimization problem Eq. [12], and the on-line estimated solution $\hat{\rho }(t)$ is the estimation of $\rho (t)$. In the experiments, the fidelity $f(t)$ is used to represent the performance of state estimation: $f(t)=Tr\sqrt{\hat{\rho }(t)^{1/2} \rho (t)\hat{\rho }(t)^{1/2} } {\kern 1pt} $. The initial state of the 1/2 spin system is $\rho (0)=[\begin{array}{cccc} {{3\mathord{\left/ {\vphantom {3 4}} \right. \kern-\nulldelimiterspace} 4} } & {{-\sqrt{3} \mathord{\left/ {\vphantom {-\sqrt{3}  4}} \right. \kern-\nulldelimiterspace} 4} ;} & {{-\sqrt{3} \mathord{\left/ {\vphantom {-\sqrt{3}  4}} \right. \kern-\nulldelimiterspace} 4} } & {{1\mathord{\left/ {\vphantom {1 4}} \right. \kern-\nulldelimiterspace} 4} } \end{array}]$, and the Bloch sphere coordinate of $\rho (0)$ is $\left({\sqrt{3} \mathord{\left/ {\vphantom {\sqrt{3}  2}} \right. \kern-\nulldelimiterspace} 2} ,{\rm \; 0,\; }{1\mathord{\left/ {\vphantom {1 2}} \right. \kern-\nulldelimiterspace} 2} \right)$. The interval time between two weak measurements is $\Delta t=0.1$ a.u., the measure efficiency is set $\eta =0.5$ and $dW=\sigma \cdot randn(2,2)$ and the variance of noise $\sigma =0.02$. The total number of estimated values under the continuous weak measurements is 200 times in the experiments. We do the experiments in two cases with different control strength $u_{x} $, interaction strength $\xi$ and the Lindblad operator $L$: Case 1: $u_{x} =0$, $\xi _{1} =0.3$, $L =\xi _{1} \sigma _{z} $; Case 2: $u_{x} =2,{\rm \; }$$\xi _{1} =0.3$, and $L =\xi _{1} \sigma _{z} $.

Fig. 3 shows the evolution trajectories of the actual state $\rho (t)$ of the quantum system and the on-line estimated state $\hat{\rho }(t)$ in the Bloch sphere under different parameters. From Fig. 3(a) one can see that in the case $u_{x} =0$, due to Lindblad operator $L =\xi _{1} \sigma _{x} $ with orthogonal to $H_{0} $, resulting the measurement operators $M_{k_{l} } $ are orthogonal to $H_{0} $, which makes the on-line estimated states are always on the z axis. When $u_{x} \ne 0$, as shown in Fig. 3(b), the measurement operators $M_{k_{l} }$ are not orthogonal to $H_{0} $, on-line estimations of the quantum state can achieve more than 95\% accuracy of fidelity.

\begin{figure}[!htbp]
    \centering
    \begin{minipage}[t]{0.2\textwidth}
       \centerline{\includegraphics[width=1.0\textwidth]{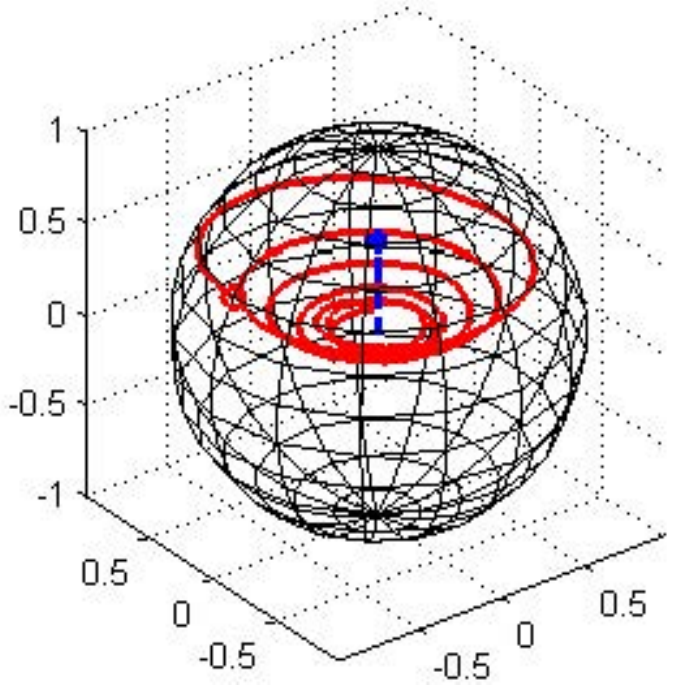}}
       \centerline{(a)}\label{fig3a.eps}
    \end{minipage}
    \hfill
    \begin{minipage}[t]{0.2\textwidth}
         \centerline{\includegraphics[width=1.0\textwidth]{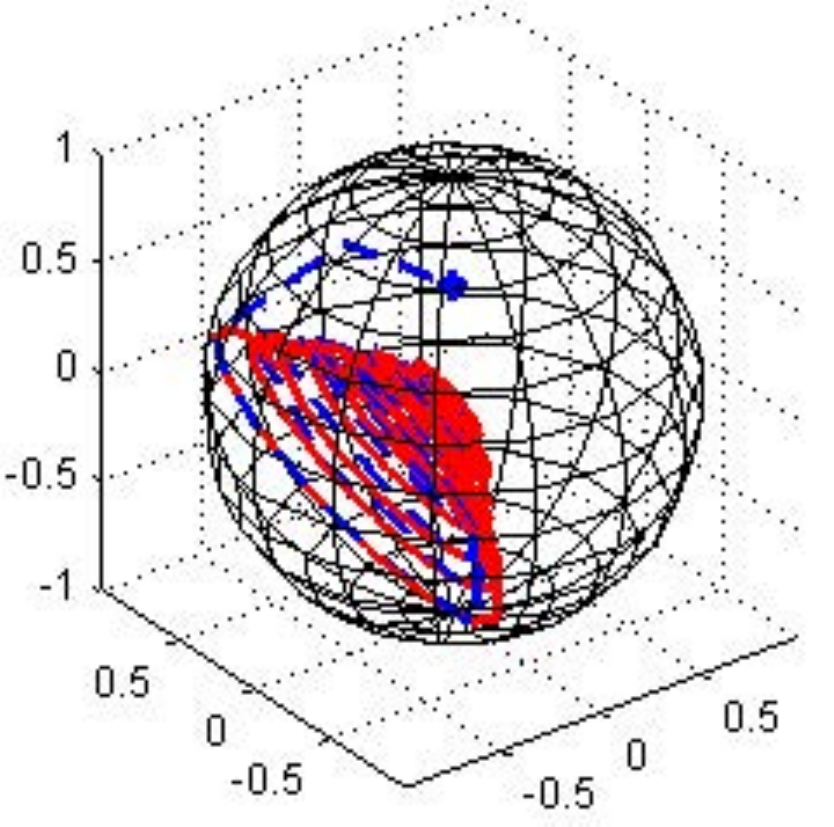}}
          \centerline{(b)}\label{fig3b.eps}
    \end{minipage}
\caption{ Evolution trajectories of the actual state $\rho (t)$ and the on-line estimation state $\tilde{\rho }(t)$ in the Bloch sphere under different parameters in the case: (a) $u_{x} =0$, $\xi _{1} =0.3$, $L =\xi _{1} \sigma _{x} $ and initial measurement operator $M_{k_{1} } (0)=\sigma _{z} $, (b) $u_{x} =2$, $\xi _{1} =0.3$, $L =\xi _{1} \sigma _{z} $ and initial measurement operator $M_{k_{1} } (0)=\sigma _{z} $;
in which the red solid line corresponds to the actual state, the blue dotted line corresponds to the on-line estimation state, "${\rm o }$" represents the position of the actual initial state $\rho (0)$, and "$*$" represents the initial of estimation state $\hat{\rho }(t)$.
}\label{fig3}
\end{figure}

\begin{figure}[!htbp]
    \centering
     \begin{minipage}[t]{0.23\textwidth}
         \centerline{\includegraphics[width=1.0\textwidth]{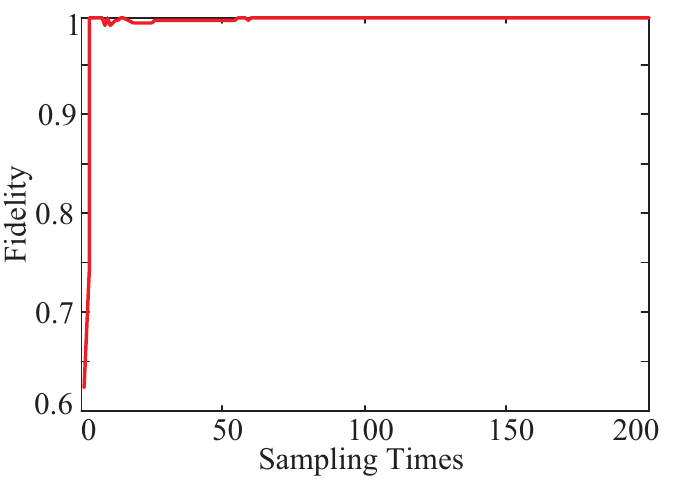}}
        \centerline{(a)}\label{fig4a}
    \end{minipage}
    \hfill
    \begin{minipage}[t]{0.22\textwidth}
         \centerline{\includegraphics[width=1.0\textwidth]{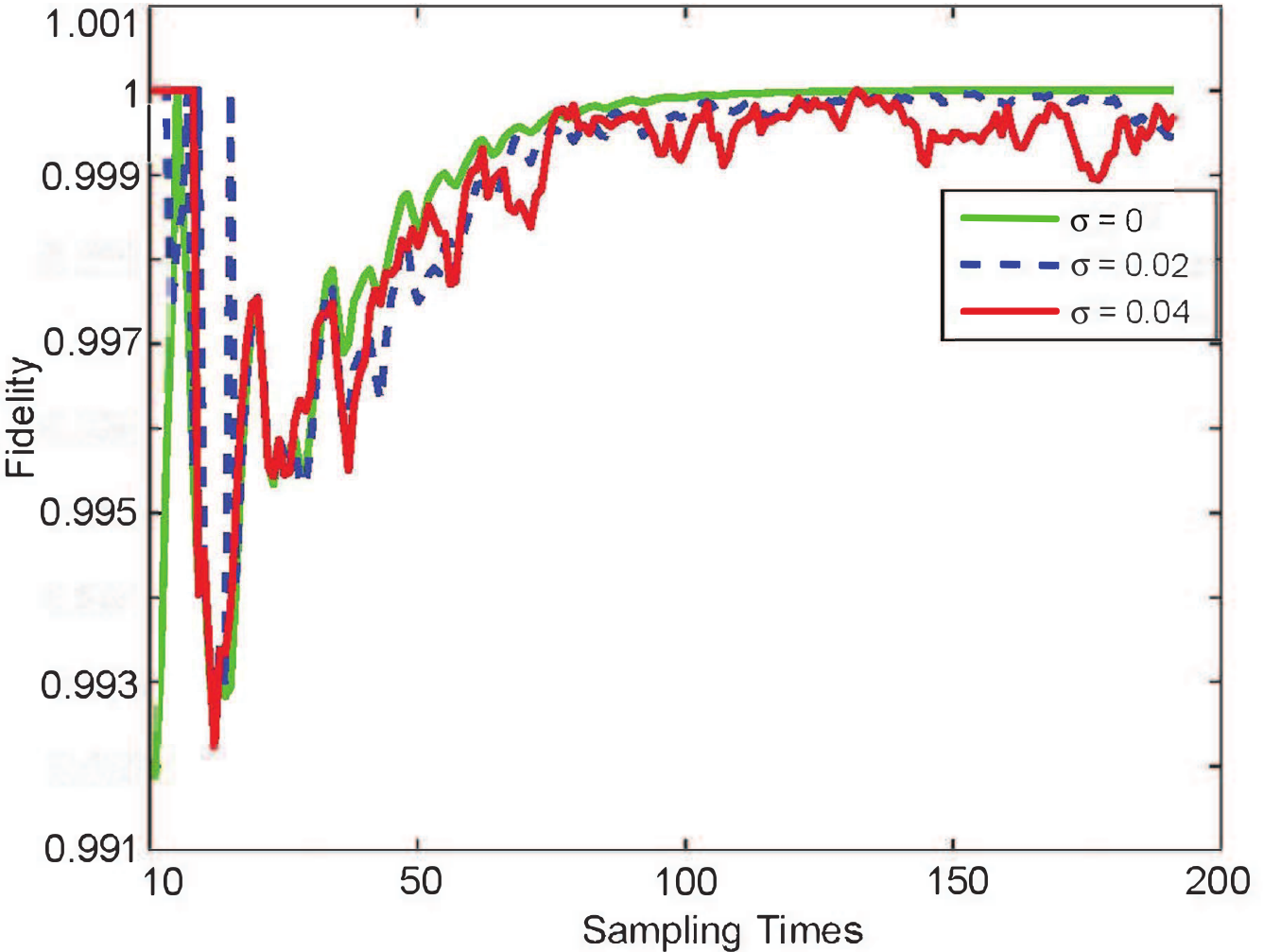}}
          \centerline{(b)}\label{fig4b}
    \end{minipage}
\caption{  Fidelity is the function of sampling times with  (a) $dW=0.02randn(2,2)$; (b) Three different noise amplitudes.}\label{fig4}
\end{figure}

From experimental results we can conclude that the Lindblad operator $L$ determines the direction of decoherence, the control strength $u_{x} $ determines the evolution direction of the state trajectory in Bloch sphere, and the interaction strength of weak measurement $\xi _{1} $ determines the speed of decoherence. The value of control strength does not affect the accuracy of estimation, and after the second measurement, all estimated states are accurate. But due to the existence of the noise, the estimated results have fluctuations. The fidelity with 200 times on-line measurement records and estimations with the variance of noise $\sigma =0.02$ is shown in Fig. 4(a). In order to investigate the effects of the noise to the fidelity, we do the experiments of the fidelity with different values of noise. The results are shown in Fig. 4(b), in which the variances of noise $\sigma $ are 0, 0.02 and 0.04, respectively.

We have proposed a new protocol for on-line quantum state estimation based on the continuous weak measurements of a dynamic evolution ensemble system. The estimation technique is nondestructive and exploits the compressed sensing theory, providing an implementable method for more complex application of high accurate closed-loop quantum feedback control. This is particular interest for microscopic systems.

This work was supported by the National Natural Science Foundation of China under grant no. 61573330 and 61720106009

\bibliographystyle{apsrev4-1} 

\end{document}